\DocumentMetadata{
  lang  = en, 
  pdfstandard = ua-2, 
  tagging = on 
}

\documentclass[sigconf]{acmart-tagged}

\AtBeginDocument{%
  \providecommand\BibTeX{{%
    \normalfont B\kern-0.5em{\scshape i\kern-0.25em b}\kern-0.8em\TeX}}}

\usepackage{tikz}
\usepackage{multirow}
\usepackage{comment}
\usepackage{float}
\usepackage{tabularx} 
\usepackage{environ}
\usepackage{xparse}
\usepackage{hyperref}
\usepackage{wrapfig} 

\usepackage{graphicx}

\usepackage{enumitem}
\usepackage{dirtytalk} 
\usepackage[inkscapelatex=false]{svg} 
\usepackage{balance}

\renewenvironment{quote}
  {\list{}{\rightmargin=0.27in \leftmargin=0.27in}%
   \item\relax}
  {\endlist}

\newenvironment{sayit}[1]
{\textit{\say{#1}}}

\copyrightyear{2026}
\acmYear{2026}
\setcopyright{cc}
\setcctype{by}
\acmConference[ASSETS '26]{The 28th International ACM SIGACCESS Conference on Computers and Accessibility}{October 25--28, 2026}{Vila Nova de Gaia, Portugal}
\acmBooktitle{The 28th International ACM SIGACCESS Conference on Computers and Accessibility (ASSETS '26), October 25--28, 2026, Vila Nova de Gaia, Portugal}
\acmDOI{10.1145/3797867.3829050}
\acmISBN{979-8-4007-2521-0/2026/10}

\begin{document}

\title{“Coder first, advocate second, college student third”: The Liminality of Going to College as a Blind Computing Student}


\author{Isabela Figueira}
\affiliation{%
  \institution{University of California, Irvine}
  \city{Irvine}
  \state{CA}
  \country{USA}
}\email{i.figueira@uci.edu}

\author{Josahandi M. Cisneros}
\affiliation{%
  \institution{University of California, Irvine}
  \city{Irvine}
  \state{CA}
  \country{USA}
}\email{josahamc@uci.edu}

\author{Stacy M. Branham}
\affiliation{%
  \institution{University of California, Irvine}
  \city{Irvine}
  \state{CA}
  \country{USA}
}\email{sbranham@uci.edu}

\renewcommand{\shorttitle}{“Coder first, advocate second, college student third”}

\begin{abstract}
Blind or low vision (BLV) students are less likely to graduate from college, particularly in computing. Prior work documents accessibility challenges in high school and college, but we lack understanding of the \textit{transition} process that produces this “leaky pipeline.” To address this, we interviewed ten BLV college students about going to college to study computing. We analyzed our data through the lens of life transition, specifically Intersecting Liminality. Our findings reveal that some BLV students face such immense digital accessibility and college acclimation barriers that the only way forward as coders is to take on a ``second job'' as a blind advocate or drop out of the computing major. We argue that the college transition is a critical point for analysis and technological intervention, and further, that Intersecting Liminality provides a useful lens for HCI scholars to unpack the compounding challenges that prevent some BLV students from completing computing degrees.
\end{abstract}






\maketitle

\section{Introduction}



As of 2024, an estimated 660,000 children in the USA have a vision disability \cite{us_census_disability_table_2024}. 
Unfortunately, while blind and low vision (BLV) students graduate high school and enter college at rates similar to sighted peers, they are significantly less likely to earn a college degree \cite{newman_post-high_2011}. Moreover, individuals with disabilities are less likely than non-disabled peers to pursue STEM and computing careers \cite{burgstahler_broadening_2007}. BLV people are thus underrepresented in the computing workforce \cite{burgstahler_broadening_2007}, perpetuating the status quo of inaccessible software tools \cite{cha_you_2024}, which has been found to negatively impact their career mobility \cite{cha_understanding_2024}. Indeed, technology barriers play a substantial role in the ``leaky pipeline'' for BLV students \cite{boadi-agyemang_understanding_2023,huff_exploring_2021,baker_educational_2019,beck-winchatz_advancing_2008,gadiraju_exploring_2021}.


Prior work in HCI has explored technology access barriers in computing classrooms in various educational contexts: high school \cite{baker_understanding_2019,huff_exploring_2021,stefik_computer_2019}, college \cite{kulkarni_case_2023,baker_educational_2019,mack_maintaining_2023}, and graduate school \cite{shinohara_burden_2021,shinohara_access_2020,shinohara_usability_2022,jain_navigating_2020,tamjeed_understanding_2021}. 
In each context, BLV students face access challenges, such as inaccessible math equations \cite{ahmetovic_latex_2021}, which necessitate time-intensive access labor and self-advocacy. 
Researchers have proposed a range of technologies to increase accessibility of Computer Science (CS) education, including physical coding languages to teach programming logic to children \cite{warren_microsoft_2019}, screen-reader-compatible plugins for mainstream Integrated Development Environments (IDEs) \cite{armaly_audiohighlight_2018, baker_structjumper_2015}, and specialized  accessible programming languages and IDEs (e.g., Quorum) \cite{stefik_accessible_2024}. 
While these works consider technology accessibility at a particular stage of a computing student's education, there has yet to be a consideration of the technology adoption process as students \textit{transition} between stages, such as transitioning from high school to college.





Life transition has been conceptualized by social scientists as a process of identity change \cite{mikal_transition_2013} comprising three stages: departure, liminal transition phase, and reincorporation \cite{van_gennep_rites_1960,haimson_social_2018}. 
During a life transition, one inhabits a liminal space, a stage between identities where one experiences ambiguity in the absence of familiar social structures \cite{turner_betwixt_1964,turner_liminality_1991}. Recently, researchers have theorized the amplified liminal experiences of disabled people undergoing multiple concurrent life transitions of aging, becoming disabled, and acquiring technology, proposing the notion of \textit{Intersecting Liminality} to describe the compounding barriers that keep people stuck in between life stages \cite{figueira_intersecting_2024, figueira_intersecting_2026}.

In this paper, we posit that going to college to study computing is one such period, comprising concurrent life transitions of becoming a college student while adopting educational and assistive technologies (AT) for a computing major. To explore the affordances of the life transition and Intersecting Liminality frameworks to reveal new intervention points for the HCI community, we pose the following research questions: 
\begin{enumerate}
    \item How do BLV students manage the transition to college computing disciplines? 
    \item What social and technical challenges do BLV students experience during this transition?
\end{enumerate}

We conducted semi-structured interviews with ten BLV participants about their experiences transitioning to college computing. 
Through an abductive thematic analysis~\cite{braun_using_2006}, we read our data through a lens of life transition and Intersecting Liminality \cite{figueira_intersecting_2026}. 
Participants described experiences reflecting Intersecting Liminality, as well as \textit{anti-structure} and \textit{communitas} \cite{turner_liminality_1991}---additional features of liminality. 
Many faced compounding barriers that kept them stuck in a cycle of adopting technology and struggling to make academic progress, hindering their transition to becoming college students. We contribute the first detailed accounts of BLV students navigating college computing transitions, along with design recommendations for what we term \textit{Transitional Education Technology}: interventions designed to support students during critical, education-focused life transitions.

\section{Related Work}

\subsection{Life Transition and Liminality}

Life transitions have been theorized as ``rites of passage'' comprising three stages: \textit{separation} from a previous identity or environment, \textit{transition}, and \textit{incorporation} into a new identity or social context \cite{van_gennep_rites_1960}. van Gennep identified this pattern underlying cultural rituals marking life changes \cite{thomassen_liminality_2014}. Turner \cite{turner_liminality_1991,turner_betwixt_1964} further theorized the transitional period of liminality, in which a person separates from social structure and enters a state of ambiguity between social structures and identities. Liminality has spatial and temporal dimensions, where one can inhabit a liminal space and experience a liminal period that can be brief or prolonged \cite{thomassen_thinking_2018}. Some people, such as older adults, have been described as existing in “persistent liminality” \cite{nicholson_living_2012}, as they perpetually balance feeling capable while navigating age-related ability loss. 

Turner argued that rites of passage produce \textit{anti-structure}, or an absence of social structure. During the liminal period, people break away from normative hierarchies that organize society and  explore different forms of social relations \cite{turner_liminality_1991, norbeck_victor_2025, turkle_living_2022}. The ``liminal persona'' exists outside of both pre- and post-transition social groups and thus does not fit into either \citep{turner_betwixt_1964}.  Occupying a space between formal roles or divisions allows for viewing social structures from new perspectives, which enables critique and transformation. Further, a person in liminality becomes a ``blank slate,'' making room for new knowledge associated with incorporation into society post-transition \citep{turner_liminality_1991}. 

As people “slip between the cracks of social structure” \cite{haggar_communitas_2024} and withdraw from usual social roles and hierarchies, they can experience a profound sense of shared humanity and connection with others in liminality---what Turner conceptualized as \textit{communitas} \cite{turner_liminality_1991}. Turner envisioned communitas as having \textit{transformative} qualities for people and society \cite{kapferer_introduction_2024}: freed from social structure, people can form new bonds and view society from a detached and critical perspective. 
Since van Gennep’s and Turner’s foundational works, liminality and communitas have become recognized as important conceptual theories \cite{thomassen_thinking_2018, thomassen_liminality_2014, szakolczai_liminality_2018} for understanding the personal and relational experiences of in-betweenness and marginality during life transition.

\subsubsection{Liminality in HCI}

Liminality is an emerging lens in HCI for understanding technology design \cite{haimson_life_2019}. Researchers have explored how social media and online communities serve as liminal spaces that support people during life transitions  \cite{semaan_transition_2016, haimson_social_2018, morioka_identity_2016}. 
Haimson~\cite{haimson_social_2018} further applied \textit{communitas} in analysis, showing that transition bloggers bond with others as they form supportive communities and ``social structure around the shared experience of gender transition'' \cite[pp. 17-18]{haimson_social_2018}.
Researchers have also examined how people  use technology during life transition rituals \cite{wolf_i_2022,kluber_designing_2020}, and how technology can support people who are isolated and unincorporated into society and its practices \cite{jensen_digital_2020}. Computing concepts (i.e., pointers and object orientation) can even hinder students from ``becoming computer scientists,'' situating them in liminal space on the threshold between novice to mastery \cite{eckerdal_limen_2007}. In Accessible Computing, this conversation is nascent, with one study and its extension conceptualizing the experiences of disabled people in \textit{Intersecting Liminality} \cite{figueira_intersecting_2024, figueira_intersecting_2026}, as they navigate multiple concurrent life transitions.

\subsubsection{Intersecting Liminality}

Building from liminality in life transition, Intersecting Liminality \cite{figueira_intersecting_2024, figueira_intersecting_2026} posits that undergoing multiple life transitions simultaneously can prolong liminality, as challenges faced during one transition can interact with and frustrate efforts to progress along another transition. In the initial study \cite{figueira_intersecting_2024}, Figueira et al. introduced the framework based on the experiences of BLV individuals as they acquire disability, acquire technology, and progress into and through older adulthood. During these transitions, some participants were trapped in liminality, unable to move forward. For example, while becoming blind and acquiring a smartphone, which has hidden and often confusing accessibility features \cite{figueira_smartphone_2023}, the absence of a blind social support network made becoming a proficient smartphone user nearly impossible. 
Figueira et al. broadened the framework in a journal extension \cite{figueira_intersecting_2026} to encompass a range technologies, disabilities, and life transitions, demonstrating that Intersecting Liminality can be fruitfully applied beyond its original context. However, accessibility research has yet to investigate the high school to college transition.  

\subsection{Accessibility of Computing in Educational Settings for BLV Students}

At each educational stage of secondary and post-secondary school, BLV students face different accessibility challenges that limit their access to essential computing course materials, tools, and curricula, affecting their academic success.

\subsubsection{College Transition}

Going to college is recognized as an important “rite of passage” into adulthood \cite{blumenkrantz_seeing_2014}, designed to support the formative experiences of emerging adults \cite{arnett_road_2023}. This transition involves separation from home, adaptation to new norms, and incorporation into academic and social communities \cite{tinto_stages_1988}. Students work to develop a sense of belonging, form friendships \cite{pittman_university_2008}, and create a ``college-centered identity'' \cite{morioka_identity_2016} as they incorporate. Many students experience challenges navigating psychological stress \cite{cooke_measuring_2006}, homesickness \cite{kelly_its_2021}, and new rigors of college academics \cite{montgomery_college_2006}. Minority and low-income students can experience greater difficulties incorporating into college culture due to disparities in preparation and resources \cite{tinto_stages_1988, venezia_transitions_2013, goldrick-rab_what_2007}. While orientation programs can serve as initiation rituals, they may not provide the extended support needed for full integration, leaving many students to navigate college life independently \cite{akli_rites_2014, tinto_stages_1988}. 

Social computing researchers have  explored technologies and interventions to support college students in adjusting to college \cite{wolf_being_2016, ha_preparing_2025, khan_navigating_2024}, including how social media in particular supports forming college identity \cite{morioka_identity_2016}, coping with homesickness \cite{kelly_its_2021}, and connecting to social supports such as with parents  \cite{smith_going_2012,deandrea_serious_2012, thomas_exploring_2022, munoz_towards_2018}. 
Accessible Computing scholars have investigated using computing technology to support skills development in young adults who are neurodivergent or have intellectual disabilities \cite{kong_understanding_2024, bayor_leveraging_2019, hong_designing_2012}, but have not yet studied how BLV young adults  navigate  adulthood or college transitions.  


\subsubsection{High School Computing for BLV Students}

Beyond assistive and mainstream technology inaccessibility in high school classrooms \cite{baker_understanding_2019, shaheen_exploring_2024}, BLV students face  barriers that restrict their participation in computing and limit early interest in the field \cite{bigham_inspiring_2008}. Both the absence of standardized computing curricula that contain accessibility topics and the inaccessibility of  educational tools further hinder BLV students’ computing exploration \cite{stefik_computer_2019, hu_diary_2021, huff_exploring_2021}. Researchers have explored several approaches to teaching programming to BLV secondary school students, including embedding accessible programming languages and IDEs into Advanced Placement Computer Science Principles (AP CSP)\footnote{Advanced Placement Computer Science Principles is an ``introductory college-level computing course that introduces students to the breadth of the field of computer science.'' \url{https://apcentral.collegeboard.org/courses/ap-computer-science-principles}} \cite{stefik_computer_2019}, virtual programming instruction \cite{hu_diary_2021}, and informal learning workshops \cite{bigham_inspiring_2008,vandegrift_game_2006, payne_different_2024}.  These initiatives aim to foster a sense of belonging and inspire students to pursue computing by providing supportive, engaging, and accessible learning environments.

\subsubsection{University Computing for BLV Students}

When BLV students go to college, they encounter a new institutional environment in which they must manage access differently, particularly within their computing majors. 
Both undergraduate and graduate students with disabilities navigate a complex, multi-stakeholder  ecosystem to access accommodations, involving disability service office (DSO) staff, professors, and others \cite{mack_maintaining_2023, tamjeed_understanding_2021}. Communication breakdowns across stakeholders occur nearly every term, often resulting in delayed or minimum accommodations \cite{mack_maintaining_2023}. In computing especially, accommodations can be ineffectual and require extra labor to make useful \cite{jain_navigating_2020, shinohara_access_2020}. 
Further, disabled students often have to discover technology access solutions independently, since DSO staff frequently lack knowledge of AT \cite{tamjeed_understanding_2021}.


One direction to improve access in CS has been to include accessibility topics in the CS curriculum itself \cite{walther_broadening_2021}. When included, accessibility has been taught mostly in HCI courses \cite{elglaly_beyond_2024, shinohara_who_2018}. Since students may not retain accessibility knowledge from a single course \cite{zhao_comparison_2020}, researchers have outlined how to create accessibility-themed assignments across multiple CS courses \cite{oleson_teaching_2026, kuang_mapping_2024, elglaly_walking_2026, el-glaly_teaching_2020, lewis_integrating_2026}. 
Notably, Computing Education research about such  interventions often has not involved disabled participants \cite{figueira_where_2026}.



While some research has focused on BLV graduate computing student experiences (e.g., \cite{jain_navigating_2020, shinohara_access_2020,shinohara_burden_2021,shinohara_usability_2022}), few studies have documented BLV undergraduate student experiences in computing. 
Existing work highlights the inaccessibility of  visually-oriented introductory computer science courses (e.g., CS1 and CS2), essential tools (e.g., IDEs, debuggers, and interface builders), and materials such as computing textbooks  \cite{baker_educational_2019, califf_helping_2008}, which can produce  discouraging learning experiences for BLV students \cite{califf_helping_2008}. 
Inaccessible engineering software and lab equipment  further exclude blind undergrads from hands-on learning \cite{kulkarni_case_2023}. Baker et al.~\cite{baker_educational_2019} specifically documented BLV programmers' coding technology and accommodation barriers, revealing social consequences of non-participation and isolation. 
Further, to our knowledge, BLV students' \textit{transitions} from high school to college computing remain underexplored in Accessible Computing research. 


\section{Methods}

\tagpdfsetup{table/header-rows={1}} 
\begin{table*}[t]
\small
\caption{Participant demographics, reported by participants. Names are pseudonyms. All participants attended a 4-year college or university.}
\begin{tabularx}{\textwidth}{
    >{\raggedright\arraybackslash}p{0.06\linewidth}
    >{\raggedright\arraybackslash}p{0.12\linewidth}
    >{\raggedright\arraybackslash}p{0.195\linewidth}
    >{\raggedright\arraybackslash}p{0.06\linewidth}
    >{\raggedright\arraybackslash}p{0.09\linewidth}
    >{\raggedright\arraybackslash}p{0.09\linewidth}
    >{\raggedright\arraybackslash}p{0.09\linewidth}
    >{\raggedright\arraybackslash}X
}
\toprule
\textbf{Name} & 
\textbf{Vision Status} & 
\textbf{Assistive Technologies Used} & 
\textbf{Gender} & 
\textbf{Race/ Ethnicity} & 
\textbf{Years in University} & 
\textbf{Switched majors?} & 
\textbf{Current Major(s)} \\
\midrule

Pardha & 
Fully Blind & 
NVDA screen reader, MathCAT & 
Man & 
Asian & 
4 (just graduated) & 
No & 
Mathematics, Computer Science \\ \hline
\addlinespace

Oliver & 
Fully Blind & 
Braille display, screen reader & 
Man & 
White or Caucasian & 
2 & 
No & 
Computer Science, Mathematics \\ \hline
\addlinespace

Walter & 
Low Vision & 
Lecture transcripts, laptop ZoomText magnifier + speech, iPhone magnifier and camera, Kurzweil ebooks & 
Man & 
White or Caucasian & 
1 & 
No & 
Cybersecurity \\ \hline
\addlinespace

Zahir & 
Visually Impaired & 
Screen reader, braille notetaker, braille display, audiobooks, tactile graphics & 
Man & 
Asian & 
5 (just graduated) & 
Out of CS & 
Business \\ \hline
\addlinespace

Thomas & Low vision. About 10\%.  Legally blind. & Large text size, magnification, sometimes a screen reader. & Man & White or Caucasian & 2 (after community college) & No & Computer Science \\ \hline
\addlinespace

Nora & 
Blind. High contrast, only larger objects and outlines & 
Braille display, talking calculator, NVDA screen reader & 
Woman & 
White or Caucasian & 
1 & 
Out of computing (Cybersecurity) & 
Transportation Safety Management \\ \hline
\addlinespace

Sahara & 
Low Vision & 
Magnifying glass, iPad to zoom in on board & 
Woman & 
Black or African-American & 
2 & 
No & 
Computer Science, Humanities \\ \hline
\addlinespace

Aiysha & 
Low Vision and Legally Blind & 
Magnification, camera, screen reader & 
Woman & 
Black or African-American & 
3 & 
Out of CS & 
Liberal Studies \\ \hline
\addlinespace

Kira & 
Closer to Blind than Low Vision. Sees large shapes but no detail & 
Screen reader primarily, sometimes uses magnification software for short inaccessible text & 
Woman & 
Native Canadian; White or Caucasian & 
1 & 
Trying to get into CS major & 
Undeclared major (Until requirements for CS are met)  \\ \hline
\addlinespace

Mei & Blind, light perception & Screen reader mostly, braille display - not all the time. & Woman & Asian & 5 (including community college) & No & Computer Science \\ 

\bottomrule
\end{tabularx}
\label{tab:demographics}
\end{table*}

\subsection{Participants}

We aimed to interview BLV students in computing majors in higher education. 
Due to blindness being a ``low-incidence'' disability and the narrowing pipeline of BLV students in  computing, we defined broad inclusion criteria. These encompassed any computing-related major or minor (e.g., computer science, cybersecurity, information technology, data science, HCI), and students who were current, prospective, or former computing majors at any stage of higher education, including those who had recently graduated. We recruited via email lists and our professional networks. 


In this study, we interviewed ten participants who identified as being blind or having low vision, had completed at least one year at a 4-year college or university in the USA or Canada, and had some contact with a computing major (ranging from previously pursued, currently pursuing, or working towards pursuing an undergraduate degree in computing). Most participants were enrolled in a college or university, and two participants had recently graduated (within a few months). See \autoref{tab:demographics} for participant demographics.




\subsection{Procedure}

We conducted semi-structured interviews with participants over Zoom, with eight interviews in  Summer 2024 and two in Fall 2025. Each interview lasted between 60 and 100 minutes (average = 80 minutes, total = 13 hours). Prospective participants completed a demographic questionnaire during recruitment and were emailed a study information sheet prior to being interviewed. Verbal consent was obtained at the start of each interview. Interviews were audio recorded and transcribed by our research team or by a professional transcription service. Participants were compensated \$40 per hour via a gift card. This study was approved by our university’s Institutional Review Board.

To explore the lived experiences of BLV students transitioning into college computing, we developed a semi-structured interview protocol grounded in life transition and Intersecting Liminality.
Interview questions spanned three categories: support systems in high school and college, preparation and planning for the transition to college, and perceived identity and belonging during transition. We asked participants to reflect back on their experiences in their first year(s) in university, asking follow-up questions to aid recall. Topics also addressed (digital) accessibility and studying computing.

\subsection{Analysis}



\begin{figure}[ht]
    \centering
    \includegraphics[width=\linewidth, alt={In the middle of a cloud is a person icon. Two axes intersect in the middle of the cloud. First axis is school, from high school student to college student. Second axis is technology, from being a technology non-user to a technology user, regarding assistive and CS major technology. Encircling the person in the middle of the liminal space cloud are two chasing arrows. On the left is a striped blue arrow pointing to the right, towards the post-transition identities. On the right is a solid red arrow pointing to the left, towards the post-transition identities. Behind the cloud is a gradient circle that is half blue (left) and half red (right), labeled sociotechnical structures.}]{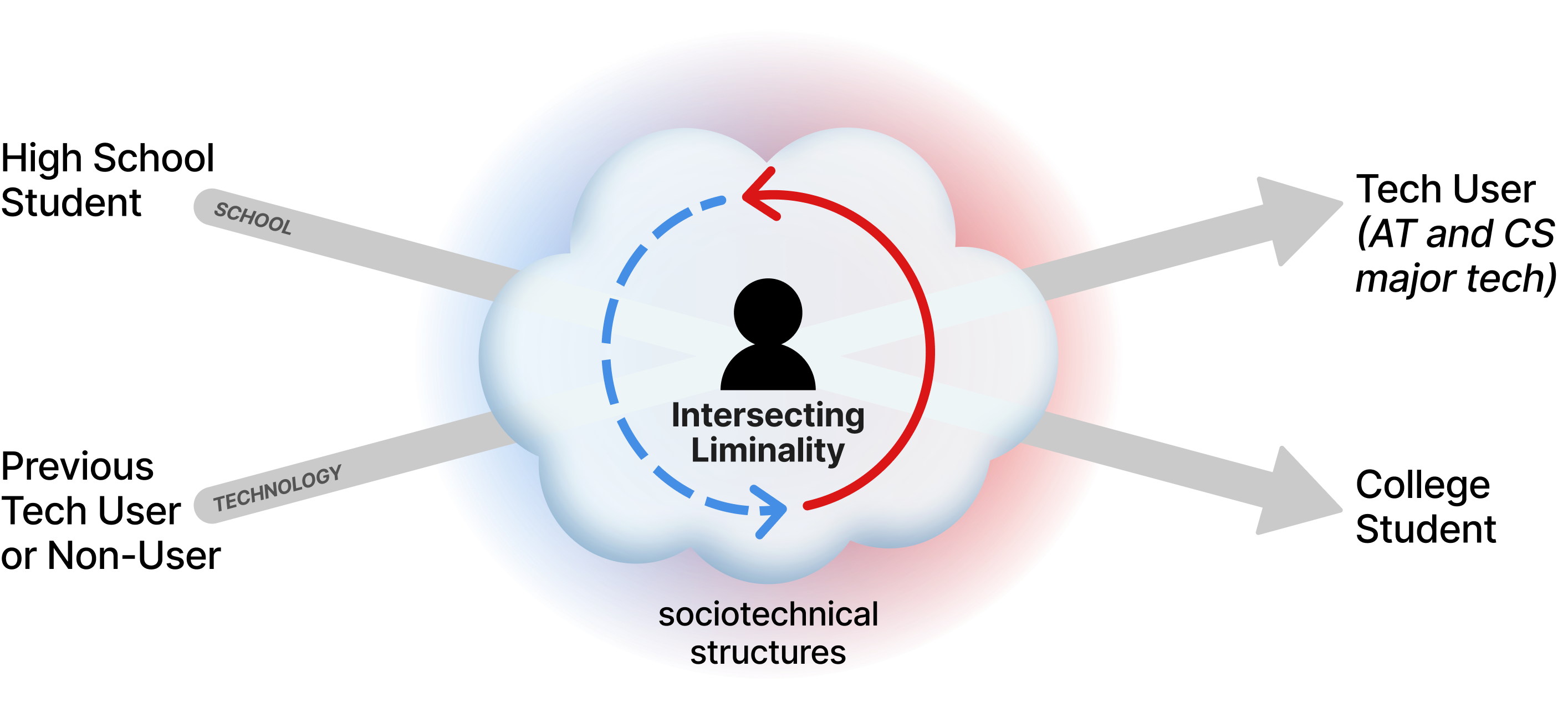}
    \caption{Intersecting Liminality. Axes of transition include ``school'' from high school to college student, and ``technology'' from non-user to user of assistive and/or CS major technology. The cloud at the transition intersection symbolizes liminality, with cyclical red and dotted blue arrows that signify sociotechnical forces keeping people locked in place. Sociotechnical context surrounds the life transition, represented with a blue and red gradient.}
    \label{fig:IL}
    \Description{Figure 1: Intersecting Liminality. In the middle of the figure there is a cloud signifying a liminal space, and in the middle of the cloud is a person icon. Two axes go through the cloud and intersect in the middle of the cloud. First axis is school, from high school student to college student. Second axis is technology, from being a technology non-user to a technology user, regarding assistive and CS major technology. Encircling the person in the middle of the liminal space cloud are two chasing arrows. On the left is a striped blue arrow pointing to the right, towards the post-transition identities. On the right is a solid red arrow pointing to the left, towards the post-transition identities. Behind the cloud is a gradient circle that is half blue (left) and half red (right), labeled sociotechnical structures.}
\end{figure}

We conducted a thematic analysis \cite{braun_using_2006} of our interview transcripts using abductive reasoning. 
We applied the analytical framework of Intersecting Liminality \cite{figueira_intersecting_2024, figueira_intersecting_2026} across two transitional axes: becoming a college student and adopting technology for studying computing (including AT and technology specific to a CS major), depicted in \autoref{fig:IL}. 
In our analysis, some participants acquired new technologies to study CS (e.g., Visual Studio Code), while others learned to use their existing AT (e.g., screen reader) in new ways to access CS curricula. As articulated in  \cite{figueira_intersecting_2024}, having to re-learn to use a technology during transition is a form of non-use to use transition. We therefore interpret both types of technology learning as existing on the non-use to use transitional axis.  
We took notice of themes related to key facets of Intersecting Liminality, such as social isolation, feelings of ``stuckness,'' becoming, and forces facilitating or hindering transition. 
Also, we enriched our analysis with Turner's constructs of \textit{anti-structure} and \textit{communitas} \cite{turner_liminality_1991}.

The first author produced initial codes organized around the phases of pre-transition, liminality, and post-transition for both transitional axes, with features of Intersecting Liminality. This author engaged in constant comparison \cite{glaser_constant_1965} across all participants' data. Initial codes and associated quotes were entered into an affinity diagram \cite{interaction_design_foundation_-_ixdf_what_2017, byrne_worked_2022} in FigJam \cite{figma_figjam_2025} and were organized into temporary groupings to search for emerging themes. Authors met regularly to discuss and reach agreement on emerging themes. 
Authors reviewed and selected final themes, which are organized temporally in this paper into the following sections: pre-college (\ref{sec:4_1}), transition period of liminality (\ref{sec:4_2_liminality}),  Intersecting Liminality examples (\ref{sec:4_3_IL}), and post-transition (\ref{sec:4_4_post}). 

\subsection{Positionality and Ethics}


Our social identities as (non)disabled, Latina/white women in computing inform our research perspectives. As women with multiple marginalized backgrounds in computing, we view the constructs of anti-structure and communitas from a personal perspective, since we have lived these in our academic computing journeys. In our meetings, we have discussed our personal experiences of in-betweenness and non-belonging as we began and completed computer science majors, and continued our careers in academia. As we share minoritized identities in computing, we appreciate resemblances between ours and our participants' stories. As we do not identify as having vision impairments ourselves, we humbly listen, notice, and probe where those experiences diverge.


Several values and philosophical frames inform our accessibility research perspectives. We are committed to an assets-based approach to disability, viewing capability of people with disabilities and not deficits. We align with the political/relational model of disability \cite{kafer_feminist_2013}, understanding that disability is generated in the environment and through interpersonal relationships, not solely within the individual. Related to this, we inspect disability through the lens of interdependence \cite{bennett_interdependence_2018}, as a dynamic production of (disabled) people and (assistive) technologies in the environment.


Our ethical procedures reflect our commitment to humanizing the people at the center of our research. We followed university IRB protocols as a baseline and made deliberate choices that went beyond those requirements. For example, in accordance with IRB, we anonymized especially identifying demographic details---replacing unique major names with broader labels such as ``Humanities'' or ``Business'' in \autoref{tab:demographics}---to protect participants while preserving data integrity. Beyond those requirements, we chose to refer to participants by pseudonyms rather than numeric identifiers, to humanize our findings and honor individual narratives.

Our commitment to a relational perspective of disability is also reflected in our research procedures. We believe that ethical research should not be extractive, but reciprocal. To that end, we actively engage with the local BLV community through hosting workshops (e.g., \cite{figueira_inspiring_2025}), providing funded scholarships, and co-creating resources and tools. We strive to treat our relationship with participants and community members as ongoing partnership.



\section{Findings}
We found that BLV students experienced Intersecting Liminality as they underwent transitions into college and into computing. First, pre-transition, we describe the degrees of access to technology, training, and support in secondary school (Section \ref{sec:4_1}). In Section \ref{sec:4_2_liminality}, we document social and technological challenges experienced while in liminality, demonstrated by social isolation and identity ambiguity (\ref{sec:4_2_1_isolation}), absence of structure (i.e., \textit{anti-structure}) such as in denied accommodations (\ref{sec:4_2_2_anti-struct}), and \textit{communitas} with other disabled students and allies (\ref{sec:4_2_3_communitas}). In Section \ref{sec:4_3_IL}, we demonstrate how two of our participants’ challenges compounded to keep them stuck in Intersecting Liminality between the two concurrent transitional axes. Finally, in Section \ref{sec:4_4_post}, we describe participants’ experiences moving forward in college, where some participants made progress and exited liminality into post-transition, and others remained in persistent Intersecting Liminality as challenges from each transitional axis impeded progress on the other.

\subsection{Before the College Transition} \label{sec:4_1}
Pre-transition to college, participants had different levels of high school support structures. Some had extensive access support and STEM skill preparation, and others had next to no structural support. 

\subsubsection{Technology Access and Training}

Most participants described having ample assistive technology support during high school from their schools and local organizations. Participants received AT such as braille displays, talking calculators, and CCTV magnifiers from school districts, plus braille displays and accessible textbooks from state disability resource offices. Many could keep their assistive technology after high school, which was \sayit{a relief not to have to get an expensive piece of assistive technology on top of everything else you need for college} (Nora). 
Most also received AT training, with several learning to use screen readers as early as middle school (e.g., Nora, Oliver). Walter attended vocational technical school for IT during high school, which helped him \sayit{adjust from doing everything on paper to doing stuff on the computer all the time} using magnification software. 

However, some students (Sahara, Aiysha) had limited access to accessible technology resources. Sahara regularly interacted with SMART boards, iPads, and laptops with  \sayit{very, very small} screens \sayit{quite literally the size of an iPad mini,} all largely incompatible with low vision.

\subsubsection{Academic and Life Skills Preparation}

Participants described varying levels of social support for learning life and academic skills to prepare for college transitions. 

Most participants (Oliver, Walter, Nora, Zahir, Kira, Mei) had extensive high school support including Individualized Education Programs (IEPs), Teachers of Visually Impaired Students (TVIs), Orientation and Mobility (O\&M) specialists, and braillists. Students also went to schools with in-house braille printing, so materials came on time. 
Kira and Mei described how braillists covered all her subjects and assignments were provided in accessible digital formats such as Microsoft Word. Kira was even permitted to use VS Code in CS class. Some students prioritized preparing for college through independent living skills, as a \sayit{main worry was moving around independently} (Zahir). Nora learned with her TVI, and Zahir attended an Adjustment to Blindness 3-month program. 

In contrast, others lacked structural support. Pardha attended high school in India, where \sayit{no such structures exist,} so his family created accessible textbooks at a local blindness resource center. Sahara became ineligible for her IEP in second grade and struggled when her low resource school \sayit{transitioned to technology.} She and her mom reached out to the district for accommodations, but were ignored. 
Sahara only discovered available state resources in college, as these resources were \sayit{never available to me in high school simply because they didn't have the right people to spread that information.}

Most participants had some CS academic preparation through STEM summer programs (Aiysha, Nora) and computing courses, though most were IT-based with little programming (Walter, Sahara, Aiysha, Nora, Zahir) or lacked a dedicated teacher (Aiysha in AP CS). A few (Kira, Oliver, Thomas) were directly exposed to programming languages in CS classes before college, while Mei was not. Math preparation varied widely from calculus and statistics (Nora, Oliver, Zahir) to remedial math (Sahara). CS and advanced math courses created expectations of college readiness that often did not match realities of computing.


\subsection{Liminality} \label{sec:4_2_liminality}
During the transition to college computing, while in liminality, students embodied ambiguous identities between computing technology users and college students (\ref{sec:4_2_1_isolation}). They also experienced the \textit{anti-structure} of lacking supports, denied accommodations, and inaccessible pedagogies (\ref{sec:4_2_2_anti-struct}). Third, they found and formed \textit{communitas} with disabled students and allies through transition (\ref{sec:4_2_3_communitas}). We also found that some students were unable to establish communitas in CS, hindering their full transition.

\subsubsection{Isolation, Both/Neither Identities} \label{sec:4_2_1_isolation}

Participants described experiences of isolation, feeling like outsiders and both/neither college CS students, indicating liminality between transitional identities. 
Some felt like such novices in CS compared to classmates, especially Sahara, who \sayit{didn’t even have access to a computer that was good enough for you and your needs until … college.} 
Nora felt unwelcome in computing clubs since no one approached her or asked, \sayit{hey, you look new here. Can I help you get started?} Rather, members showed off coding skills,  
creating an uninviting atmosphere for novices: \sayit{keep up or catch up or just, you know, don't come.} 
Aiysha felt uniquely singular in tech conferences and classes: \sayit{No one gets it. I've never really met anyone in similar shoes than me ... disabilities in tech, and in college, and of color, and not in the correct major.}

In several instances, going with or without accommodations further increased isolation in computing classes. 
In Pardha’s introductory CS class, he received alternative assignments to the visual grid language that \sayit{were not exactly the same.} Thus, it \sayit{cause[d] an early separation between my learning experience and that of a sighted counterpart.} 

A few students (Sahara, Kira, Aiysha, Mei) felt so overburdened with advocating for their access needs, that, while they technically were college students, they did not feel as such. 
Kira felt her constant self-advocacy was \sayit{a class of its own} on top of her coursework, diminishing her identity as a college student. 
Like Kira, Mei felt she \sayit{spent so much time figuring this [accessibility] stuff out and still not getting anywhere, while ... everyone else is having a good time. I don't feel like I got to enjoy the typical uni life.} 
Sahara was a \sayit{college student officially} but \sayit{holistically, I don't really think of myself as a college student.}  
She equated her chaotic college experience to \sayit{a FEMA\footnote{Federal Emergency Management Agency \url{https://www.fema.gov}} camp after a natural disaster. Everything’s in a rush. … and all you’re hoping for is that the next day will be a little bit better than the last.} Participants felt stuck navigating inaccessible structures that contradicted  college life expectations.

\subsubsection{Anti-Structure} \label{sec:4_2_2_anti-struct}

Students all encountered an absence of structure in college while in liminality. This manifested through a literal absence of structured support and students’ interactions with college structures that failed them, such as inaccessible technologies and unsupportive institutional systems that resulted in ineffectual and denied accommodations. Students perceived additional institutional dysfunctionality specific to BLV computing students, as Zahir shared, \sayit{the logistical support system … was really helpful for folks who are going into business and a non-STEM field.}

\paragraph{Denied Accommodations as Absence of Structure}
Compared to high school support structures where materials were brailled and provided in ample time, transitioning to college meant interacting with centralized disability service offices (DSOs) where staff often lacked knowledge about blindness and resources for accessible materials, leading participants to failing classes. In hindsight, participants like Kira and Mei felt that their high schools were \sayit{almost a little too accessible. That it hit me even harder getting out of it because I was not prepared} (Mei) for the level of inaccessibility in college.

While Walter’s DSO had a blindness specialist, participants regularly encountered non-supportive DSO staff while seeking alternative formats for inaccessible PDFs and books. Kira’s DSO had only two staff members, and \sayit{they knew nothing about blindness specifically} or AT. When Kira explained her math textbook was a scanned PDF, staff suggested a braille note-taker, not knowing it could not read images. Similarly, Mei's DSO outsourced transcription to campus librarians who \sayit{were having trouble understanding what is alt text,} producing slide PDFs where computing diagrams were labeled only as \sayit{the word \say{picture,}} leaving Mei to instead seek usable descriptions from friends and AI in an absence of institutional structural support. 
Further, Kira’s and Zahir’s DSOs also lacked on-campus braille printers, compared to their well-equipped high schools. At one point, Kira's DSO even suggested she pay \$60,000 for a campus braille embosser. 
When negotiating calculus accommodations for example, Kira and Zahir were effectively denied accommodations due to major delays receiving braille math materials: \sayit{since they were outsourcing stuff, … assignment packets and things would show up past the due dates of assignments, and so I'd end up falling behind} (Zahir). 
Lacking accessible materials, despite attempts to self-teach (Kira), both fell \sayit{further and further behind} (Kira), eventually failing their calculus courses. 

Some students (Nora, Sahara) found themselves reconfiguring how they receive accommodations in a web of new stakeholders, getting tangled between high school's proactive support and college's self-advocacy expectations. Nora entered college expecting the support and access she had in high school. Although Nora's TVI prepared her for requesting accommodations and  communicating with professors, she fell into a support gap, despite communicating with the DSO over summer and having a \sayit{supportive} computing professor. In the first week, Nora identified needing slides and code in advance, since following live coding lectures was \sayit{overwhelming} when \sayit{listening to the professor and screen reader.} However, VS Code's inaccessible debugging features overshadowed her C++ learning. Support fell away on every front: the course’s no-collaboration policy cut her off from peer support; she avoided the DSO since \sayit{the policy ... was that way for every student;} she prioritized accessibility troubleshooting with TAs rather than learning;
and her blind friends did not know C++. As a novice unable to imagine alternatives like more accessible IDEs, she waited for accommodations that never came.

\paragraph{Rigid Pedagogies as Absence of Structure}
Several participants illustrate how an absence of accessible pedagogical structures prevented them from progressing in their CS majors. 

First, the way that class was structured reduced the efficacy of assistive technology accommodations. For example, Nora struggled to follow along in lecture where she needed to attend to slides and real-time coding with a screen reader. 
Similarly, Sahara struggled with 30 pages of nightly reading + videos for self-learning, which she described as  \sayit{incompatible} with low vision and hands-on learning preferences. 

Second, the absence of structure in course policies around technology use further excluded BLV students from their classrooms.  Kira’s english professor barred her from using her computer to access in-classroom assignments. Kira recounted, \sayit{No matter what I said, like I even told her like, \say{I want to be successful in your class. I can’t do that without my computer.} She goes like, \say{That’s just the way my class is structured.}} The professor’s workaround excluded Kira from class: \sayit{If it’s a group assignment, get someone else to write. But if it’s a one-person assignment, don’t come to class. You’re doing it at home.} This \sayit{confusing} and \sayit{weird} (Kira) arrangement isolated her from classmates, group work, and the classroom.

Many students found themselves negotiating their access around pedagogical structures. For example, 
Oliver spent \sayit{four or five hours assuring} his calculus professor that he could in fact do math: \sayit{[The professor] was like, \say{Well, you can't see polar coordinates.} And I'm like, \say{I don't need to see polar coordinates. It's just R and theta measured out from the X axis. That's it.}}
This places an unreasonable burden on students who, as novices at the beginning of their college transitions, may not yet know the content well enough to argue for how they can access it. 

\paragraph{Math Material Conversion as Absence of Structure}

Math emerged as a critical site of anti-structure within the computing transition, since math is a requirement for computing majors that became inaccessible in college. Participants reported challenges securing accessible math materials in a timely fashion or at all, after losing dependable high school supports. 
Kira found that \sayit{math is the biggest thing taking away from my ability to get through college.} Oliver argued, \sayit{Programming is perfectly accessible. It's the math where people die.} 

On top of braille printing delays, a major issue is that \sayit{the state of the art for digital math access is poor} (Pardha), with digital materials such as PDFs not rendering math equations at all or accessibly. Oliver explained, \sayit{Screen readers can't see that line [above the radicand]. ... they see the square root symbol. They say square root. But where does it stop?} Consequently, Oliver’s \sayit{algorithms book was a no-go} due to ambiguous, complex math notations. While \sayit{MathML kind of solves this} plain text issue on websites with a navigable tree structure, most \sayit{books are still a major, major, major pain} (Oliver). Due to inaccessible math materials, Mei \sayit{had to rely on [the tutor] telling me what the heck's going on. And there was no other materials, and everything is pretty much verbal.} She passed these courses, but she felt like she learned nothing.

In high school, TVIs such as Oliver’s and Mei's provided timely accessible math materials (e.g., braille, tactile graphics) that disappeared in college, leaving students to source materials independently through their DSO or themselves.  Since braille conversion for textbooks could take \sayit{six to eight months} (Oliver), Oliver proactively contacted his DSO in June to prepare. However, the DSO accepted his disability paperwork but would \sayit{refuse to talk to you until you are really set in your course schedule,} which only occurs during orientation in July. 
In the summer between high school and college, students simultaneously  \textit{were} admitted and registered but were \textit{not yet} fully college students. As Oliver's DSO would only support full college students, this anti-structure prevented him from gaining access to the lengthy braille calculus textbook in time for classes to start. 


\paragraph{Creating Structure for Math Access}


As a result of anti-structure of math access, students typically stumbled onto potential solutions at inopportune times during difficult math classes. Oliver learned about LaTeX through \sayit{hearsay} as a braille math alternative, discovering it the summer before college after learning about impending braille material delays. Mei discovered mid-semester in a statistics course that coding math in R was much more accessible than visual diagrams: \sayit{As soon as I found out that I can do some kind of math in code, I was like, that's a lot easier than drawing diagrams, let's do it.} Rather than being introduced to these approaches proactively, students encountered them on their own, often already behind.

Adopting these workarounds required self-teaching on top of already demanding coursework, and not all were feasible. Oliver self-taught LaTeX over the summer and, after sustained effort, grew to prefer it over braille and MathML for its linear navigability with a screen reader. Kira encountered LaTeX after failing calculus due to delayed braille materials, receiving a LaTeX-format textbook from her professor before retaking the course. 
Kira had serviceable knowledge of the format: \sayit{I know enough that it’s functional. I know, like, ‘frac’ represents a fraction and, like, the basic operations. But I’m not, like, proficient. But I’m sure that will come with time.} However, learning a new math representation while learning calculus, as someone used to reading math in braille, might impose an additional challenge.
Mei recognized LaTeX as an option, but unlike Kira, did not pursue learning it:  \sayit{There's not an easy way to learn it then ... and it's hard to read.}  She relied instead on her tutor as a scribe. She passed her math courses but felt she had learned little.

For students who did adopt LaTeX, securing it as an official accommodation required additional advocacy labor. 
Oliver described the LaTeX format as a \sayit{nonstandard} accommodation that he had to \sayit{push hard for} with his DSO. He labored to secure LaTeX materials from multiple stakeholders: exams and book chapters transcribed by his DSO, homework assignments directly from CS professors who maintained LaTeX materials, and even textbooks solicited directly from authors.

In the absence of structure in math, students created structure: they discover workarounds by chance, then adopt them through self-teaching, and finally advocate to secure institutional recognition for accommodations.

\subsubsection{Communitas} \label{sec:4_2_3_communitas}
Participants described bonding with other students and becoming advocates for themselves and others undergoing similar transitions. Forming communitas with other disabled students and allies supported participants in making transitional progress, while isolation from community further kept participants  stuck in liminality.

\paragraph{Forming Communitas}
Some participants formed communitas on campus, through bonding with college friends which increased their sense of belonging as a college student. Sahara found a circle of friends \sayit{who are willing to support me and not necessarily see my visual impairment as a crutch for making excuse.} 
Outside of her university, Kira luckily \sayit{did find an online community of blind people to talk to} on Discord that helped her feel understood and were \sayit{extremely instrumental} to her activism. She shared that they were \sayit{the first people that triggered my big uprising with advocacy, because eventually they told me it wasn’t normal, … not all schools were like that.} This communitas supported her \sayit{advocacy journey} during her transition to college computing. 


Several participants experienced communitas as a  process into advocacy identity, demonstrating the personally and structurally \textit{transformative} properties of liminality and communitas. Sahara turned \sayit{the ableism and the discrimination}  into \sayit{my fuel for advocacy.} Kira practically became an activist as she labored to educate her DSO about blindness and AT, spending countless hours emailing unresponsive DSO staff, and enlisting \sayit{another blind person who knew about rights} to help. The DSO started \sayit{taking things seriously} when she mentioned her human rights lawyer. Kira felt her work would make structural changes:  \sayit{By the time I’m done with the school, they’ll be [in] tiptop accessibility shape.} Mei described \sayit{a great consequence} of inadequate institutional support: \sayit{Now I know the whole thing in my life is going to be somewhat around accessibility and getting more people into STEM. ... I don't want this to happen again to someone else.} 
Oliver channeled his experience into creating an online \sayit{resource list … for the community} of blind CS students, distributing the institutional knowledge he had labored to find alone. Through communitas, these students produced change in their personal identities and in the inaccessible structures around them. 

\paragraph{Isolation from Communitas}
However, several participants mentioned feeling isolated from spaces where they may have found support, barring them from the potential to form communitas through typical collegiate bonding. Since Kira was not an official CS major, she felt unable to join the campus computing club: \sayit{There’s the computer science student association. ... it might be social, might not. It sounds like you have to be a computer science student … in order to get in.} Mei felt she lacked the time to join computing clubs since it \sayit{feels more like extra effort I have to put in when I'm already so busy.} 

Others felt isolated from blind peers studying computing. Mei also felt unable to talk about her challenges gaining access in her CS major with her blind friends since they were not STEM majors: 
\sayit{the other blind students were just like, \say{just do something else then.} 
She felt like she was unable to talk about her challenges gaining access in her major \sayit{mostly because too much of it were technical and so it was hard to explain.}} Nora expected to lean on her blind friends from high school who were CS or math majors  when learning programming in college, since they often would \sayit{pick each other's brains whenever we have a question.} However, since her friends knew Python and not C++, she was alone: 
\begin{quote}
    \sayit{I was kind of thrown right into C++. I didn't really have anyone from the blind community to kind of be like, `hey, how would you do this?' ... I wasn't able to get any advice from them. I was kind of figuring this out on my own.} (Nora)
\end{quote}
Zahir similarly felt program-L\footnote{\url{https://www.freelists.org/archive/program-l/}}, a blind programming online community, was unapproachable since its topics were too \sayit{ultra specific} and professionally advanced to engage with as a student. 
Participants thus learned programming alone, isolated from other blind computing students and potential mentors.

\subsection{Intersecting Liminality} \label{sec:4_3_IL}
We observed compounding challenges that kept students \textit{stuck} between transitions, that is Intersecting Liminality, in four participants’ experiences (Kira, Nora, Sahara, and Aiysha). To demonstrate how Intersecting Liminality can differentially manifest, we explore Nora's and Kira’s stories. 

\subsubsection{Nora's Isolation from Supports}

Nora faced the challenge of learning new technology for computing (tech axis), while being isolated from her blind friends and sighted classmates (college axis). She ultimately failed to find adequate communitas to navigate the anti-structure produced by inadequate accommodations and classroom structures. 

Nora entered college expecting to lean on peer support of her blind friends from high school and new friends from college. However, she felt \sayit{thrown right into C++,} isolating her from blind peers who only knew Python. She also found it impossible to form a new support network in her major, since her computing professor did not allow student collaboration. Neither could she find support in computing clubs, since Nora felt excluded by students who already knew \sayit{five programming languages} and were eager to show off.

Lacking peer support, Nora felt forced to \sayit{figure things out alone}: \sayit{being new to it [C++] and then being blind on top of that} meant she \sayit{had to learn what everyone else needed to learn} regarding computing concepts, \sayit{and I had to learn how would I do this [with a screen reader].} Further, desiring to just be a \say{normal} college student, she felt unable to unlock additional accommodations from TAs or the DSO to succeed in her computing class: 
\begin{quote}
    \sayit{[My question for TAs] was more of a visual \say{where is this highlighting?} Not a, \say{hey, can you write my code for me?} Kind of thing. I was careful to keep it as close to what a normal student would get help with.} (Nora)
\end{quote}
In other words, Nora's desire to be a \sayit{normal college student}---a student who adheres to sighted environment norms conflicting with her access needs---prevented her from asking for the accommodations she needed in computing classes. Thus, she decided to leave her computing major to prioritize her college transition.

\subsubsection{Kira's Advocacy Burden}

Kira’s advocacy for accessible materials and AT  (technology axis), prevented her from making adequate academic progress to declare a computing major (college axis). Though Kira found communitas, the extreme anti-structure of establishing appropriate accommodations, she transformed into an advocate at the expense of becoming a college student and computer science major.

Like Nora, Kira had a great deal of access support in high school. She believed that her high school \sayit{set me up to fail, because they gave me all the extra accommodations.} Thus, when she encountered unknowledgeable DSO staff, Kira realized she was ill-prepared \sayit{for all the accessibility curve balls} she would soon face in college. 

To overcome accessibility barriers in college, Kira started a \sayit{big uprising} in advocacy labor to access materials and AT in her classes. For example, in Calculus class, Kira spent months advocating for tactile graphics and braille materials, with DSO staff who \sayit{flaked off,} only to receive materials too late to pass the class. In English class, Kira spent hours trying to convince her professor to allow her to user her AT during class.
Advocating for access was so much work that Kira had to take on a lighter course load: 
\begin{quote}
    \sayit{I feel like right now, I have to take two [classes] just in case one of them blows up in my face and I need to advocate. Like, trying to juggle advocacy with two classes was a lot. ... advocacy is basically a class of its own.} (Kira)
\end{quote}
She was left with little capacity to simply be a college student: 
\begin{quote}
    \sayit{I feel like I am less of a college student and more of an advocate, to be honest, right now. I feel like I’m more so just there because it’s what you’ve got to do to get into coding. So I’m a coder at heart. \textbf{I’m a coder first, ... advocate second, college student third.}} (Kira)
\end{quote} 
Though still a coder at heart, the full-time job of advocacy is stalling her college transition on multiple fronts: taking classes at a slower rate, re-taking failed classes at a higher rate, and ultimately unable to meet requirements to transition out of her Undeclared major into CS.

\subsection{Moving Forward in College} \label{sec:4_4_post}
While some students persisted in Intersecting Liminality, others forged paths forward, prioritizing their computing majors or their broader college experience.

\subsubsection{Moving through Liminality}
Most students described making progress as a computing student or leaving their computing majors, which we viewed as moving through liminality. 

Oliver, Pardha, Thomas, and Walter described making progress and feeling belonging in their CS majors while adapting to college. Pardha, who graduated and is pursuing a CS Master's degree, felt that hard computing courses increased his self-confidence. Oliver, Thomas, and Walter all felt belonging in computing, with Walter sharing, \sayit{I really like my major … I think it’s cool to show other students ... everybody can do computing in some way just maybe a little differently.} 
Thomas described his progress in college as \sayit{circling the drain in a good way. There’s clearly some kind of goal that I’m working towards.} While Walter was still searching for a close friend group, all four students felt adjusted to college and demonstrated confidence in their disability identity. 

Both Nora and Zahir progressed through liminality by choosing to exit their CS majors. In her first college semester, Nora foresaw misery after attending unwelcoming CS clubs, saying, \sayit{If I have to work with people like this for the rest of my life, I'm not going to be happy.} Nora left CS entirely, finding her new classmates \sayit{more down to earth.} Zahir, who enjoyed programming but struggled in STEM classes,  \sayit{realized that we only have one opportunity in college to do what we're passionate about,} graduating instead with a business/entrepreneurship degree while maintaining a path toward a tech career. Both felt adapted to college. 

\subsubsection{Persistent Intersecting Liminality}
In four cases (Kira, Sahara, Aiysha, Mei), we observed persistent Intersecting Liminality, akin to persistent liminality \cite{nicholson_living_2012}, in which participants remained stuck in transitions to college and in technology adoption for computing majors. 


Two participants felt stuck attempting to adapt and reorient course material and technology use to access computing, while advocating for access needs. 
Sahara was desperate to make progress in her computing major, repeatedly retaking classes, struggling to find workable solutions.
This summer, Sahara was on her third attempt retaking an introductory Python after taking an incomplete, and she was eager to retake a failed electronics class after her professor acquired accessible software. 
Mei felt \sayit{too far in [in CS] to start something ... completely unrelated from scratch,} and she chose to study computing at another university online after three inaccessible in-person semesters.

Kira and Aiysha both experienced access challenges---respectively in math course material access and remote accommodation options---keeping them stuck outside their intended computing majors. Both felt unlike college students due to constant self-advocacy work and inability to pursue computing majors. Both decided to independently learn computing. Kira self-taught LaTeX, Swift, HTML, CSS, JavaScript, and Python, \sayit{to be ahead of the game.} She also had a \sayit{blind coding mentor} who shared accessible online coding resources. 
Aiysha was repeatedly denied remote attendance accommodations for psychosocial disability, and she was forced to drop her IT major despite several attempts to return. Instead, she will graduate with a liberal studies degree online, planning to 
\sayit{get my degree and then have all of my skills and certifications outside of that.} Despite persistent liminality, both remained determined to reach computing careers. 
\section{Discussion}


While prior HCI research has focused on technology accessibility in isolated educational contexts, none have yet examined the  \textit{transition} between high school and college computing,  a ``critical juncture'' for disabled students \cite{burgstahler_accessstem_2011}. 
Through our analysis of BLV computing students’ experiences of going to college, we apply life transition theory—and in particular the concepts of anti-structure, communitas, and Intersecting Liminality—to this domain for the first time, exposing new phenomena and potential intervention sites that context-isolated studies cannot identify. Our data show compelling evidence that BLV students navigated liminality as they encountered anti-structure and formed communitas. Students who simultaneously underwent a college transition and a technology transition faced compounding challenges, leaving some stuck in persistent Intersecting Liminality. 
In this education transition context, our findings demonstrate Intersecting Liminality’s utility as an analytic lens in Accessible Computing and HCI. We propose Transitional Education Technology as an umbrella term for a novel class of technology interventions designed to reduce anti-structure and cultivate communitas for students in transition.

\subsection{Why Transitions are Consequential for Accessible Computing Education} 

By applying the lens of life transition to our data, we identified both previously documented and novel hindrances to obtaining computing degrees.  
Like prior work, we found that BLV students struggled with inaccessible assignments, software, and digital books in college \cite{baker_educational_2019, kulkarni_case_2023, califf_helping_2008}. Further, BLV computing students navigated a complex multi-stakeholder ``accessibility ecosystem'' \cite{mack_maintaining_2023, ly_accessibility_2025} to access accommodations, often resulting in communication breakdowns with unknowledgeable DSO staff \cite{tamjeed_understanding_2021, mack_maintaining_2023}, delayed receipt of accessible materials \cite{califf_helping_2008, mack_maintaining_2023}, and accommodations that proved ineffectual \cite{jain_navigating_2020, shinohara_access_2020}. 


In the absence of access structures, we noticed participants had to learn new technologies and re-learn to use familiar technologies in order to gain access in their computing major. For example, Oliver was unable to access his college math textbooks, despite requesting braille materials the summer before courses began. As a result, he had to identify his own LaTeX math workaround, teach himself the LaTeX code, and advocate for this \sayit{nonstandard} technology accommodation with his DSO. While learning how to code and how to debug, Nora had to re-learn how to use her screen reader to access VS Code---reacquiring her AT for this new domain. This nuance is in line with the existing Intersecting Liminality model \cite{figueira_intersecting_2024, figueira_intersecting_2026}, since people undergoing multiple life transitions have been found to have to reacquire familiar technologies to gain access.

Additionally, our analysis revealed challenges at the \textit{intersection} of the college and computing major technology transitions. For example, our participants had to navigate the tension between wanting to feel like a ``normal'' college student and needing to confidently self-advocate for disability accommodations to succeed academically. This aligns with prior findings that BLV college students are concerned about fitting in \cite{califf_helping_2008}, and BLV K-12 students avoid technology use that differentiates them from sighted peers \cite{ahmetovic_latex_2021, baker_understanding_2019}. Our interpretation posits that students whose disabled identities conflicted with sighted norms or expectations risked forgoing accommodations entirely (Nora), while those who did advocate often became so overwhelmed by the process that they had little time or energy left to simply be a college student (Sahara, Kira, Aiysha). 
Consequently, some BLV students became stuck in Intersecting Liminality, as the weight of unresolved disability identity and accessibility barriers compounded to impede progress. Others exited liminality, but did so at the cost of leaving the computing major altogether.

Notably, the four participants who we interpreted as persisting in Intersecting Liminality identified as women of color. Research has shown that underrepresented students in CS, such as women and people of color, face challenges including low representation, limited mentorship \cite{kricorian_factors_2020,robnett_research_2018, lopez_beyond_2026}, inadequate support \cite{morrow_inclusion_2026}, and under-recognition \cite{carlone_understanding_2007} that hinder progress.  From a perspective of intersectionality \cite{crenshaw_mapping_1991}, which describes how overlapping marginalized identities produce unique forms of oppression \cite{crenshaw_mapping_1991}, research has documented how people with multiple marginalized identities experience compounding challenges that hinder success in CS \cite{rankin_intersectional_2020, blaser_perspectives_2020}. Our findings of compounding marginality arose through a temporal analysis of identity during life transition(s) of people with disabilities. This resonates with a broader body of HCI and Accessible Computing literature calling for greater attention to intersectionality, including class, race/ethnicity, gender, and disability \cite{desai_race_2026, bennett_its_2021, schlesinger_intersectional_2017, kumar_intersectional_2019}. More research is needed to understand the broader systemic injustices facing disabled people with additional marginalized identities in computing.

\subsection{Reflecting on Intersecting Liminality as an Analytic Lens} 

We argue that an analytic lens of Intersecting Liminality \cite{figueira_intersecting_2024, figueira_intersecting_2026}, augmented with anti-structure and communitas (this paper), helps us identify novel phenomena and locate novel sites for intervention. \textit{Liminality} surfaces the identity tensions and social isolation, \textit{anti-structure} exposes the structural absences between institutions where support disappears, and \textit{communitas} identifies the peer connection and shared identity during transition that students need to move through liminality. \textit{Intersecting Liminality} mobilizes these concepts across multiple transitions to guide interventions that might facilitate the journey through liminality, avoid persistent liminality, and minimize accessibility-related exits from computing majors. 


One such site for intervention is before the college transition begins. 
To reduce the technology adoption burden in college, transition planning for high school students could incorporate hands-on experience with discipline-specific technologies: accessible mainstream IDEs and programming languages, digital math formats, and college lab equipment. Self-advocacy instruction and materials explaining how college accommodations work may further prepare students to secure the supports they need once on campus.  Finally, connecting students with national networks of peers who share the same disability and career aspirations might jumpstart communitas development.

Another site for intervention is during the liminal phase, including the time after high school, leading up to college, and as college begins. Universities might explore formal pre-enrollment processes---such as early accommodation setup and structured technology orientation---to support admitted students like Oliver in the summer before classes begin. Tools such as an accessibility dashboard in Learning Management Systems (LMSs) \cite{patatas_real-time_2026} or similar systems already in use could educate high school teachers and professors about the ``accessibility ecosystems'' \cite{mack_maintaining_2023} in both contexts. As another example of a potential intervention, workshops like ``Accessing STEM in Higher Education'' \cite{figueira_inspiring_2025} that include K-12 and college BLV students, high school teachers, and university professors can build communitas and surface gaps between accessibility norms in secondary and post-secondary STEM education. 

Beyond these intervention sites, transition support should recognize that students may simultaneously navigate multiple life transitions or other major life events that compound with the challenges of academic transition. 
More broadly, Intersecting Liminality holds promise as a lens for examining transitions into and out of other contexts actively studied in HCI---such as graduate school \cite{jain_navigating_2020, shinohara_access_2020, tang_access_2026},  workplace \cite{das_it_2019, cha_understanding_2024, cha_you_2024}, home \cite{nogueira_putting_2025}, family roles changes such as becoming a care partner \cite{johnson_i_2025} or parent \cite{durrant_admixed_2018, cassidy_cuddling_2024, chheda-kothary_engaging_2024}, and disability onset \cite{baker_understanding_2019, motahar_exploring_2024}---to explore the relational, social, and technological challenges that can bar people from progressing toward their life goals. 

\subsection{Transitional Education Technology}

Despite a growing body of work demonstrating how transitions between academic institutions are critical for students with disabilities (e.g.,  \cite{jain_navigating_2020, tang_access_2026, burgstahler_accessstem_2011}), both with respect to (assistive) technology adoption and maturation as a student, the technology supports available in the education domain (Educational Technology) offer little built-in support for transition. We propose that life transition and Intersecting Liminality can provide a useful frame for a wide range of academic transitions (e.g., between grades, between campuses, etc.) and across disability identities. 
We operationalize this as \textit{Transitional Education Technology}: a design frame for developing technology that meets students where they are in a transition, equipping them for the technology demands ahead while considering the supports they are leaving behind. 
We explore three directions below to illustrate this frame.



\subsubsection{Technology Support for New Student Orientation}


At most universities, incoming students attend new student orientation programs. Separately, BLV students often leverage a variety of transition planning programs often documented in a transition plan required within their Individualized Education Programs (IEPs) \cite{IDEA_transitions}: including vocational training programs like Pre-Employment Transition Services (Pre-ETS) \cite{Perkins_understanding_2026, lambert_perspectives_2023, adiani_career_2022}, Orientation \& Mobility (O\&M) training to navigate physical campus spaces \cite{Braille_inst_orientation_2026, HKSB_orientation_2026}, and AT training with a TVI or AT specialist to navigate braille displays and screen readers \cite{shaheen_exploring_2024, Braille_inst_services_2026}. However, in college, these access supports disappear, and no equivalent exists for navigating the computing major technology landscape. \textit{Technology orientation} could fill this gap and alleviate some college transition pressure, promoting independent learning of assistive and educational technology by taking on the functions of an AT specialist or TVI when students lose access to prior supports.

Such orientation could come in the forms of a tutorial plug-in for a mainstream IDE or a smart voice- or text-based conversational tutor, orienting BLV computing students to using AT while learning computing concepts. Such tools, often already in use in college classrooms (e.g., custom LLM for learning in AP CSP \cite{frazier_customizing_2024}), could help students like Nora learn to navigate C++ code and a debugger within the VS Code IDE. 

Technology orientation could also be embedded in existing institutional structures on both sides of the transition. High school technology training (specific to STEM majors as well as AT) could be incorporated into IEP transition planning or delivered through programs like Pre-ETS that already support students preparing for transitioning out of secondary school. College STEM departments could include discipline-specific technology orientations during general college orientation or in a dedicated introductory course, to introduce students to tools central to the major.

\subsubsection{Digital Mathematics for College STEM Access}



Strikingly, several participants independently converged on computing itself as a math access strategy, pointing to a design opportunity unique to the CS major. Mei found coding math in R more accessible than navigating visual diagrams; Kira proposed solving calculus limits with algorithms; and Oliver sought out LaTeX math because of its linear, code-like structure, which parallels programming conventions, making it an accessible format for BLV students already developing programming fluency. However, participants discovered these workarounds by chance. Learning a new math format mid-transition was laborious, involving searching for materials and a steep learning curve that could impede adoption. Preparing earlier during the transition could reduce the adoption labor when taking more advanced courses in a computing major. 

Despite our observation of digital math as an access strategy, little research attends to teaching digital math equation notation. Existing work on math access for BLV students has largely focused on Nemeth braille \cite{herzberg_experiences_2023}, tactile graphs \cite{phutane_tactile_2022,smith_role_2012, rosenblum_teachers_2018, zebehazy_charting_2014, baker_tactile_2016, chen_tactile_2025, ramoa_leveraging_2025, giudice_learning_2012}, 
virtual graphs \cite{ohshiro_making_2021, sharif_voxlens_2022, kim_math_2019}, 
and digital equation editors \cite{ge_stereomath_2024}. 
Despite LaTeX’s prevalence in higher education \cite{maneki_latex_2012, huang_access_2025}, only one study in Accessible Computing explored teaching LaTeX in high school science classrooms \cite{ahmetovic_latex_2021}. 
The inconsistent and untimely discoveries of digital math  in college makes it a compelling direction for Transitional Education Technology research. 
Future work should examine whether this gap in digital math preparation is widespread among BLV students, and if so, what structural or pedagogical barriers prevent earlier adoption.


\subsubsection{Online Communities for Communitas}

Finally, online communities could serve as transitional technology that supports BLV computing students in forming communitas and moving through liminality. 
Existing networks for blind STEM students (including AccessSTEM \cite{do-it_accessstem_2026}, AccessComputing \cite{accesscomputing_accesscomputing_2026}, and NFB mailing lists \cite{NFB_nfbnetorgs_2026}) face challenges of visibility, scale, and specificity that point to a design opportunity for an online community organized around stage of college computing transitions. Students could identify their current transitional stage and connect with peers and mentors, filtering by transitional stage, disability, institution, and major. Thus, an incoming disabled student could locate a more senior disabled student at their institution who  knows which IDE certain CS courses use, which DSO staff member understands AT needs, and which professors are worth contacting early about accommodations. 

Rather than a dedicated social platform, social elements could be embedded in educational technologies such as smart tutors or conversational agents (e.g., for peer support \cite{liu_compeer_2024}), IDEs, or LMSs, increasing student connections on platforms already broadly in use. By reducing the barrier to finding peers and mentors at critical transitional moments, such communities could serve as the mechanism through which BLV computing students form the communitas they need to support their advocacy efforts, move through liminality, and make progress in computing.

\section{Limitations}

Our study was conducted in the United States. Participants attended 4-year colleges or universities across the USA and Canada. Stages of education varied across participants, from first-year college students to recent graduates. The differences in time since beginning college transition may have affected participant recall and reflections.  We recruited  participants from a narrow sector of the population: college students tend to fall within a narrow age bracket, and we studied computing majors. Thus, our findings regarding the education system and challenges experienced may not apply to other contexts. Future work should compare challenges transitioning across additional academic contexts such as trade schools and community colleges, and also to the workplace.

\section{Conclusion}

We investigated the experiences of BLV students transitioning into college computing, examining how college and CS major technology life transitions intersect and hinder progress. While all participants navigated liminality, we found that some BLV students experienced \textit{Intersecting Liminality}, becoming stuck between transitions as absence of structure and social support hindered their progress toward computing degrees. 
Some technologies, such as digital math, proved to be both a potential transition barrier and an opportunity to support the accessible college computing transition.  We propose \textit{Transitional Education Technology} as a design frame for developing interventions to support students undergoing critical educational and technology transitions.





\bibliographystyle{ACM-Reference-Format}
\bibliography{BLVtransitions-references, extrarefs}





\end{document}